\documentclass[twocolumn]{jpsj2} %% two-column layout
\title{New method to simulate quantum interference using deterministic processes
and application to event-based simulation of quantum computation%
}

\author{H. \textsc{De Raedt}$^{1}$\thanks{E-mail address: deraedt@phys.rug.nl; http://www.comphys.org},
K. \textsc{De Raedt}$^{2}$\thanks{E-mail address: deraedt@cs.rug.nl}
and K. \textsc{Michielsen}$^{1}$
\thanks{E-mail address: kristel@phys.rug.nl}}

\inst{$^{1}$Department of Applied Physics, Materials Science Centre, University of Groningen,\\
Nijenborgh 4, NL-9747 AG Groningen, The Netherlands\\
$^{2}$Department of Computer Science, University of Groningen,\\
Blauwborgje 3, NL-9747 AC Groningen, The Netherlands}

\abst{We demonstrate that networks of locally connected
processing units with a primitive learning capability
exhibit behavior that is usually only attributed to quantum systems.
We describe networks that simulate
single-photon beam-splitter and Mach-Zehnder interferometer experiments
on a causal, event-by-event basis and demonstrate that
the simulation results are in excellent agreement with quantum theory.
We also show that this approach can be generalized to simulate
universal quantum computers.}
\kword{computer simulation, quantum theory, quantum interference,
quantum computation, machine learning}

\begin{document}
\maketitle

\def\ORDER#1{\hbox{${\cal O}(#1)$}}
\def\BRA#1{\langle #1 \vert}
\def\KET#1{\vert #1 \rangle}
\def\EXPECT#1{\langle #1 \rangle}
\def\BRACKET#1#2{\langle #1 \vert #2 \rangle}
\def\hbar{{\mathchar'26\mskip-9muh}}
\def\mod{{\mathop{\hbox{mod}}}}
\def\CNOT{{\mathop{\hbox{CNOT}}}}
\def\Tr{{\mathop{\hbox{Tr}}}}
\def\bPsi{{\mathbf{\Psi}}}
\def\bPhi{{\mathbf{\Phi}}}
\def\bzero{{\mathbf{0}}}
\def\Eq#1{(\ref{#1})}
\def\NOBAR#1{#1}
\def\BAR#1{\overline{#1}}
\def\openone{\leavevmode\hbox{\small1\kern-3.8pt\normalsize1}}

\def\DLM{DLM}
\def\DLMS{DLMs}

\section{Introduction} %% No sections necessary for express letters, letters and short notes
Computer simulation is a powerful methodology to model physical phenomena~\cite{LAND00}.
However, some of the most fundamental experiments in quantum physics~\cite{GRAN86,TONO98}
have not been simulated in the event-by-event manner in which the experimental observations
are actually recorded~\cite{PerfectExperiments}.
In experiments the detection of events appears to be random~\cite{GRAN86,TONO98},
in a sense which, as far as we know, has not been studied systematically.
Quantum theory gives us a recipe to compute the frequency of
the observation of events but it does not describe individual events, such as
the arrival of a single electron at a particular position 
on the detection screen~\cite{TONO98,HOME97,FEYN65,BALL03}.
Reconciling the mathematical formalism (that does not describe single events) with
the experimental fact that each observation yields a definite
outcome is often referred to as the quantum measurement paradox.
This is a central, fundamental problem in the foundation
of quantum theory~\cite{FEYN65,HOME97,PENR90}.
Therefore, it is not a such a surprise that
within the framework of quantum theory, no algorithm has been found to perform
an event-based simulation of quantum phenomena.

From a computational viewpoint, quantum theory provides us
with a set of rules (algorithms) to compute
probability distributions~\cite{HOME97,KAMP88,QuantumTheory}.
Therefore we may wonder what kind of algorithm(s) we need to perform an event-based
simulation of the experiments~\cite{GRAN86,TONO98} mentioned above without using
the machinery of quantum theory.
Evidently, the present formulation rules out any method based on the solution of the
(time-dependent) Schr{\"o}dinger equation
and we have to step outside the framework that quantum theory provides.

In this paper we demonstrate that locally-connected networks of processing units
with a primitive learning capability are sufficient
to simulate deterministically and event-by-event,
the single-photon beam splitter and Mach-Zehnder interferometer
experiments of Grangier et al.~\cite{GRAN86}.
We also show that this approach can be generalized to simulate
universal quantum computation by a deterministic event-by-event process.
Thus, the method we propose can simulate wave interference phenomena
and many-body quantum systems using classical, particle-like processes only.

Our results suggest that we may have discovered
a procedure to simulate quantum phenomena
using causal, local, deterministic and event-based processes.
Our approach is not an extension of quantum theory in any sense
and is not a proposal for another interpretation of quantum mechanics.
The probability distributions of quantum theory are
generated by a deterministic, causal learning process,
and not vice versa~\cite{PENR90}.

\setlength{\unitlength}{1cm}
\begin{figure*}[t]
\begin{center}
\includegraphics[width=14cm]{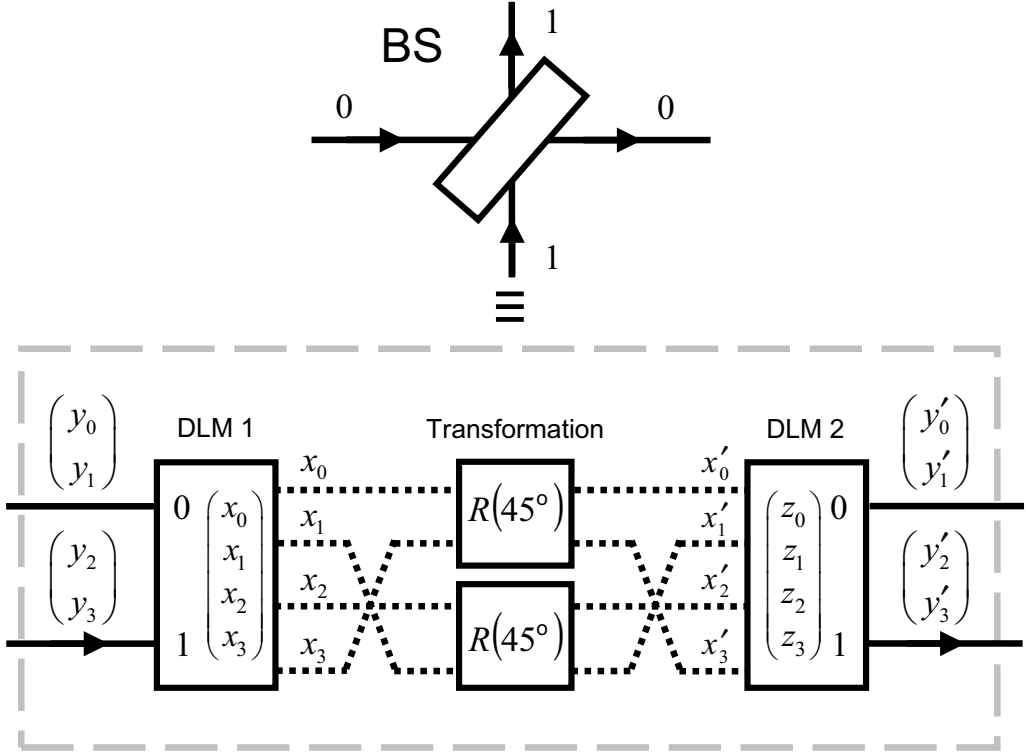}
\caption{
Diagram of the network of two \DLMS\ that performs a deterministic
simulation of a single-photon beam splitter (BS) on an event-by-event basis~\cite{MZIdemo}.
The solid lines represent the input and output channels of the BS.
Dashed lines indicate the flow of data within the BS.
Input channel 0 receives $(y_{0},y_{1})=(\cos\psi_0,\sin\psi_0)$
with probability $p_0$.
Input channel 1 receives $(y_{2},y_{3})=(\cos\psi_1,\sin\psi_1)$
with probability $p_1=1-p_0$.
$R(45^\circ)$ denotes a rotation of a two-dimensional
vector by $45^\circ$ (see Eq.~(\ref{Rphi})).
}
\label{dlms}
\label{figbs}
\end{center}
\end{figure*}

\section{Deterministic Learning Machine (DLM)~\cite{MZIdemo}}\label{sec2}
In quantum physics, an event corresponds to the detection of a photon, electron,
and the like.
In our simulation approach an event is the arrival of a message
at the input channel of a processing unit.
This processing unit typically contains two \DLMS\ (described below).
We use the diagram of a \DLM-based processor that performs
the event-by-event simulation of single-photon beam splitter,
as shown in Fig.~\ref{dlms}, to describe the operation
of the different components of the processor.
The applications to quantum computations presented later
demonstrate that the structure of the \DLM-based processor
is in fact generic.

In Fig.~\ref{dlms}, the presence of a message is
indicated by an arrow on the corresponding line.
The first component, \DLM\ 1, ``learns'' about the occurrence of
an event on one of its two input channels that we label with 0 and 1.
For brevity, we refer to an event on channel 0 (1) as a 0 (1) event.
The second component transforms the data stored in \DLM\ 1
and feeds the result into \DLM\ 2.
\DLM\ 2 ``learns'' this data.
Finally the learning process itself is used to determine whether
\DLM\ 2 responds to the input event by sending out either a 0 or a 1 event.
None of these components makes use of random numbers,
hence the name deterministic learning machine.

Usually, \DLM-based simulation algorithms contain
several \DLM-based processors that form a network.
In this paper we only consider networks of processing units in which
only one message is traveling through the network at any time.
Thus, the network receives an event at one of its inputs, processes
the event and delivers the processed message through one of its output channels.
After delivering this message the network can accept a new input event.

\subsection{Description of a \DLM}

A \DLM\ is a very simple, classical dynamical system with a primitive learning capability.
This dynamical system consists of a unit vector
(such as ${\bf x}=(x_{0},x_{1},x_{2},x_{3})$ in \DLM\ 1 of Fig.~\ref{dlms}),
a rule that specifies how this vector changes when an input event is received,
and a rule by which the \DLM\ determines the type of output event
it generates as a response to the input event.
The initial value of the internal vectors is irrelevant.
In simulations, we usually use random numbers to initialize the internal vectors
of all the \DLMS\ in the network.

We now describe the learning process of a \DLM\ in detail~\cite{KOEN04}.
The basic idea of the learning algorithm is that
the \DLM\ minimizes the distance between the input vector (discussed later)
and the internal vector and that this minimization is sufficient
to construct particle-like processes that mimic quantum phenomena.
However, the algorithm that we describe below
cannot be derived from the axioms of quantum theory.
After many trials and failures, we simply discovered that learning algorithms
of this type can be used to simulate quantum phenomena.

First we consider \DLM\ 1 in Fig.~\ref{dlms}.
The internal state of \DLM\ 1 is represented by
the vector ${\bf x}=(x_{0},x_{1},x_{2},x_{3})$.
\DLM\ 1 can accept two different types of input events,
but only one at a time.
Event 0 carries a message
represented by a two-dimensional unit vector $(y_{0},y_{1})$.
Event 1 carries a message
represented by a two-dimensional unit vector $(y_{2},y_{3})$.
Upon receiving an input event, \DLM\ 1 performs the following steps:
%
%\begin{enumerate}
\begin{itemize}
\item
\DLM\ 1 computes eight candidate internal states
\begin{eqnarray}
{\bf w}_1&=&(+\sqrt{1-\alpha^2+\alpha^2x_0^2},\alpha x_{1},\alpha x_{2},\alpha x_{3}),\nonumber \\
{\bf w}_2&=&(-\sqrt{1-\alpha^2+\alpha^2x_0^2},\alpha x_{1},\alpha x_{2},\alpha x_{3}),\nonumber \\
{\bf w}_3&=&(\alpha x_{0},+\sqrt{1-\alpha^2+\alpha^2x_1^2},\alpha x_{2},\alpha x_{3}),\nonumber \\
{\bf w}_4&=&(\alpha x_{0},-\sqrt{1-\alpha^2+\alpha^2x_1^2},\alpha x_{2},\alpha x_{3}),\nonumber \\
{\bf w}_5&=&(\alpha x_{0},\alpha x_{1},+\sqrt{1-\alpha^2+\alpha^2x_2^2},\alpha x_{3}),\nonumber \\
{\bf w}_6&=&(\alpha x_{0},\alpha x_{1},-\sqrt{1-\alpha^2+\alpha^2x_2^2},\alpha x_{3}),\nonumber \\
{\bf w}_7&=&(\alpha x_{0},\alpha x_{1},\alpha x_{2},+\sqrt{1-\alpha^2 +\alpha^2x_3^2}),\nonumber \\
{\bf w}_8&=&(\alpha x_{0},\alpha x_{1},\alpha x_{2},-\sqrt{1-\alpha^2 +\alpha^2x_3^2}).
\label{HYP4}
\end{eqnarray}
The parameter $0<\alpha<1$ controls the learning process
and is discussed in more detail later.
The plus and minus sign in
front of the square roots is introduced to allow the vector ${\bf w}_j$
to cover the whole eight-dimensional unit sphere.
\item
If \DLM\ 1 receives an input event of type 0 with message $(y_{0},y_{1})$,
it constructs a vector ${\bf \hat x}=(y_{0},y_{1},x_{2},x_{3})$.
If \DLM\ 1 receives an input event of type 1 with message $(y_{2},y_{3})$,
it constructs a vector ${\bf \hat x}=(x_{0},x_{1},y_{2},y_{3})$.
\DLM\ 1 determines the update rule $m$ that minimizes the cost function
\begin{equation}
C_j = -{\bf w}_j^T{\bf \hat x},
\label{HYP3}
\end{equation}
that is, $C_{m}\le C_{j}$ for $j=1,\ldots,8$.
\item
\DLM\ 1 updates its internal vector by replacing
${\bf x}$ by ${\bf w}_m$.
\item
\DLM\ 1 generates a new (internal) event by putting
the values of its internal vector
on its four output channels.
\item
\DLM\ 1 waits for the arrival of the next input event.
%\end{enumerate}
\end{itemize}

The transformation stage applies an orthogonal transformation $T$
to ${\bf x}=(x_{0},x_{1},x_{2},x_{3})$.
In general, the precise form of the transformation $T$ depends
on the particular function that the processor has to perform.
In the example shown in Fig.~\ref{dlms}, the orthogonal transformation $T$
takes two pairs of elements from ${\bf x}$ and performs the plane rotation
\begin{eqnarray}
R(\phi)=\left(
\begin{array}{cc}
\cos\phi&-\sin\phi\\
\sin\phi&\phantom{-}\cos\phi
\end{array}
\right)
,
\label{Rphi}
\end{eqnarray}
with $\phi=\pi/4$. As we show later, this transformation implements
the single-photon beam splitter.
The result ${\bf x^\prime}=(x_{0}^\prime,x_{1}^\prime,x_{2}^\prime,x_{3}^\prime)$
of this transformation is sent to the input of \DLM\ 2.
Thus, \DLM\ 2 accepts messages in the form of a four-dimensional unit vector.
\DLM\ 2 updates its internal vector ${\bf z}=(z_{0},z_{1},z_{2},z_{3})$
according to the following procedure:
\begin{itemize}
\item \DLM\ 2 performs computes eight candidate internal states
\begin{eqnarray}
{\bf w}_1&=&(+\sqrt{1-\alpha^2+\alpha^2z_0^2},\alpha z_{1},\alpha z_{2},\alpha z_{3})
,\nonumber \\
{\bf w}_2&=&(-\sqrt{1-\alpha^2+\alpha^2z_0^2},\alpha z_{1},\alpha z_{2},\alpha z_{3}),\nonumber \\
{\bf w}_3&=&(\alpha z_{0},+\sqrt{1-\alpha^2+\alpha^2z_1^2},\alpha z_{2},\alpha z_{3}),\nonumber \\
{\bf w}_4&=&(\alpha z_{0},-\sqrt{1-\alpha^2+\alpha^2z_1^2},\alpha z_{2},\alpha z_{3}),\nonumber \\
{\bf w}_5&=&(\alpha z_{0},\alpha z_{1},+\sqrt{1-\alpha^2+\alpha^2z_2^2},\alpha z_{3}),\nonumber \\
{\bf w}_6&=&(\alpha z_{0},\alpha z_{1},-\sqrt{1-\alpha^2+\alpha^2z_2^2},\alpha z_{3}),\nonumber \\
{\bf w}_7&=&(\alpha z_{0},\alpha z_{1},\alpha z_{2},+\sqrt{1-\alpha^2 +\alpha^2z_3^2}),\nonumber \\
{\bf w}_8&=&(\alpha z_{0},\alpha z_{1},\alpha z_{2},-\sqrt{1-\alpha^2 +\alpha^2z_3^2}).
\label{HYP2}
\end{eqnarray}
\item \DLM\ 2 determines the update rule $m$ that minimizes the cost function
\begin{equation}
C_j = -{\bf w}_j^T{\bf x}^\prime,
\label{HYP1}
\end{equation}
that is, $C_{m}\le C_{j}$ for $j=1,\ldots,8$.
\item \DLM\ 2 updates its internal vector by replacing
${\bf z}$ by ${\bf w}_m$.
\item \DLM\ 2 generates an output event of type 0 (1)
if $m=1,\ldots,4$ $(5,\ldots,8)$, carrying
the message $(y_{0}^\prime,y_{1}^\prime)=(z_{0},z_{1})$
($(y_{0}^\prime,y_{1}^\prime)=(z_{2},z_{3})$).
\item \DLM\ 2 waits for the arrival of the next input event.
\end{itemize}
Comparing the algorithms for \DLM\ 1 and \DLM\ 2, we see that
they are indentical except for part of the second step and the fourth
step in which the output is generated.

\subsection{Dynamic behavior of a \DLM}

In general, the behavior of a \DLM\ defined by rules
Eqs.~\Eq{HYP4} and \Eq{HYP3} or
Eqs.~\Eq{HYP2} and \Eq{HYP1} is difficult to analyze without the use of a computer.
However, for a fixed input ${\bf x}^\prime={\bf u}$, it is clear what a \DLM\ will do.
It will minimize the cost given by Eq.~\Eq{HYP1} by rotating
its internal vector ${\bf z}$ to bring it as close as possible to ${\bf u}$.
After a number of events (depending on the initial value of
${\bf z}$, the input ${\bf u}$, and $\alpha$),
${\bf z}$ will be close to ${\bf u}$.
However, the vector ${\bf z}$ does not converge to a limiting value
because the \DLM\ always changes its internal vector state by a nonzero amount.
It is not difficult to see (and supported by simulations, results not shown)
that once ${\bf z}$ is close to ${\bf u}$,
it will keep oscillating about ${\bf u}$~\cite{KOEN04}.
Below we analyse this behavior in more detail using \DLM\ 2 as an example.
The dynamics of \DLM\ 1 is the same as that of \DLM\ 2.

Let us denote by $n_0$ the number of times the
\DLM\ selects update rule $m=1,2$ (see Eq.\Eq{HYP2}).
Writing
\begin{eqnarray}
w^2_{0,m}=1-\alpha^2+\alpha^2 z_0^2\equiv (z_0+\delta)^2,
\end{eqnarray}
and assuming that $0\ll\alpha<1$,
we find that the variable $z_0$ changes by an amount
\begin{eqnarray}
\delta\approx(1-\alpha^2)(1-z_0^2)/2z_0,
\end{eqnarray}
where we have neglected terms of order $\delta^2$.
Similarily, if $N$ is the total number of events then $N-n_0$ is the number of times
the \DLM\ selects update rules $m\not=1,2$.
For $j\not=1,2$, Eq.\Eq{HYP2} gives
\begin{eqnarray}
w^2_{0,j}=\alpha^2 z_0^2\equiv (z_0+\delta^\prime)^2,
\end{eqnarray}
where we have neglected terms of order ${\delta^\prime}^2$.
Hence $z_0$ changes by
\begin{eqnarray}
\delta^\prime\approx-(1-\alpha^2)z_0/2.
\end{eqnarray}

If ${\bf z}$ oscillates about ${\bf u}$
then $z_0$ also oscillates about ${u}_0$.
This implies that
the number of times $z_0$ increases times the increment must approximately be equal
to the number of times $z_0$ decreases times the decrement.
In other words, we must have $n_0\delta+(N-n_0)\delta^\prime\approx0$.
As $z_0\approx{u}_0$ we conclude that $n_0/N\approx {u}_0^2$.
Applying the same reasoning for the cases where
the \DLM\ selects update rule $m=3,4$ shows that
the number of times the \DLM\ will apply
update rules $m=3,4$ is proportional to ${u}_0^2+{u}_1^2$.

At this point, there is not yet a  relation
between the dynamics of the \DLM\ and quantum theory.
However, let us now assume that
$p_0=z_{0}^2+z_{1}^2$ ($p_1=z_{2}^2+z_{3}^2$)
is the probability that a quantum system
is observed to be in the state 0 (1).
In quantum theory, we would describe this state
by a wave function with complex amplitudes
$\hat z_{0}+i\hat z_{1}$ ($\hat z_{2}+i\hat z_{3}$).
Let is now consider a \DLM\ that is learning the
four values ${\bf z}=(z_{0},z_{1},z_{2},z_{3})$.
From the foregoing discussion, it follows that
once the \DLM\ has reached the stationary state in
which it oscillates about ${\bf z}$,
the rate at which the \DLM\ uses update rules
$m=1,2,3,4$ ($m=5,6,7,8$) corresponds to the probability
$p_0$ ($p_1$)
to observe a 0 (1) event in the quantum mechanical system.
Thus, the \DLM\ generates 0 and 1 events in a deterministic manner and
because it generates 0 (1) events if it selected update rule $m=1,2,3,4$ $(5,6,7,8)$,
the rate at which these events are generated
agrees with the corresponding probabilities of quantum theory.
As the applications presented later demonstrate,
this correspondence is all that is needed to perform
an event-by-event simulation of quantum interference
and many-body quantum phenomena.

\begin{figure*}[t]
\begin{center}
\includegraphics[width=14cm]{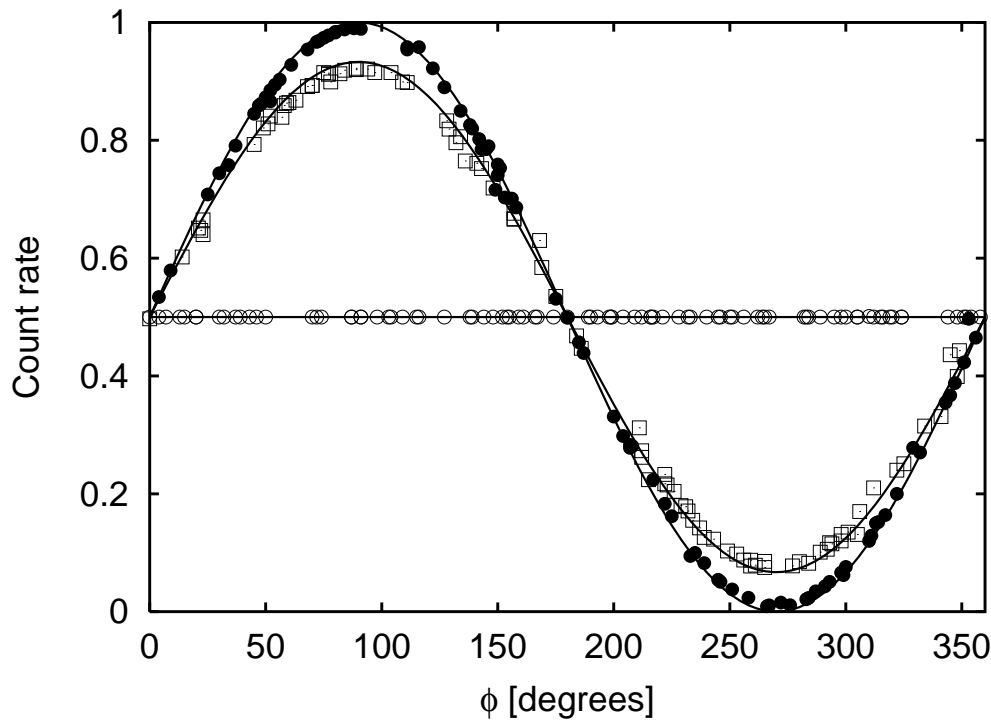}
\caption{
Simulation results for the beam splitter shown in Fig.~\ref{figbs}.
Markers give the simulation results for the
normalized intensity $N_0/(N_0+N_1)$
in output channel 0 as a function of
$\phi=\psi_0-\psi_1$.
Open circles: $p_0=1$;
Bullets: $p_0=0.5$;
Open squares: $p_0=0.25$.
The results of quantum theory $|b_0|^2$ (see Eq.~(\ref{B0}))
are represented by lines.
}
\label{one-bs}
\end{center}
\end{figure*}

\subsection{Stochastic variant}

The sequence of events that is generated by a \DLM\ (network) is strictly deterministic
but a simple modification turns a \DLM\ into a stochastic learning machine (SLM).
The term {\sl stochastic} does not refer to the learning process but to
the method that is used to select the output channel that will carry the outgoing message.
As explained earlier, in the stationary regime
$x_0^2+x_1^2$ and $x_2^2+x_3^2$
(or $z_0^2+z_1^2$ and $z_2^2+z_3^2$)
correspond to the probabilities of quantum theory.
Thus, a comparision of for instance, $x_0^2+x_1^2$,
with a uniform random number $0<r<1$ gives the probability for sending the message
over the corresponding output channel.
Although the learning process of this processor is still
deterministic, in the stationary regime the output events are
randomly distributed over the two possibilities.
Of course, the frequencies of output events
is the same as that of the original \DLM-network.
Replacing \DLMS\ by SLMs in a \DLM-network changes the order in which
messages are being processed by the network but leaves the content of the
messages intact.

\subsection{Generalization}

In the previous discussion, we considered a
\DLM-based processor (see Fig.~\ref{dlms}) that accepts
two different types of events whereby each event
carries a message containing two real numbers.
This is sufficient to simulate quantum phenomena such as
single-photon interference but if we would like to perform
event-by-event simulations of more complicated
quantum systems such as quantum computers,
a generalization is necessary.
From the foregoing description of the
learning rule of a \DLM\, it is obvious
how this rule may be generalized to handle
an arbitrary number $N_e$ of different
events of messages of arbitrary (but of the same) length $N_m$:
Use a vector of $N_e N_m$ elements to represent
the internal state of the \DLM\ and,
instead of eight candidate rules, compare the cost
of $2N_e N_m$ candidate rules.
Clearly, the construction of \DLM-based networks is
very systematic and straightforward.

\subsection{Summary}

A \DLM\ responds to the input event by
choosing from all possible alternatives, the internal state
that minimizes the error between the input and the internal state itself.
This deterministic decision process is used to determine
which type of event will be generated by the \DLM.
The message contains information about the decision the \DLM\ took
while updating its internal state and, depending on the application,
also contains other data that the \DLM\ can provide.
By updating its internal state, the \DLM\ ``learns" about the input events it receives
and by generating new events carrying messages, it tells
its environment about what it has learned.

\begin{figure*}[t]
\begin{center}
\includegraphics[width=12cm]{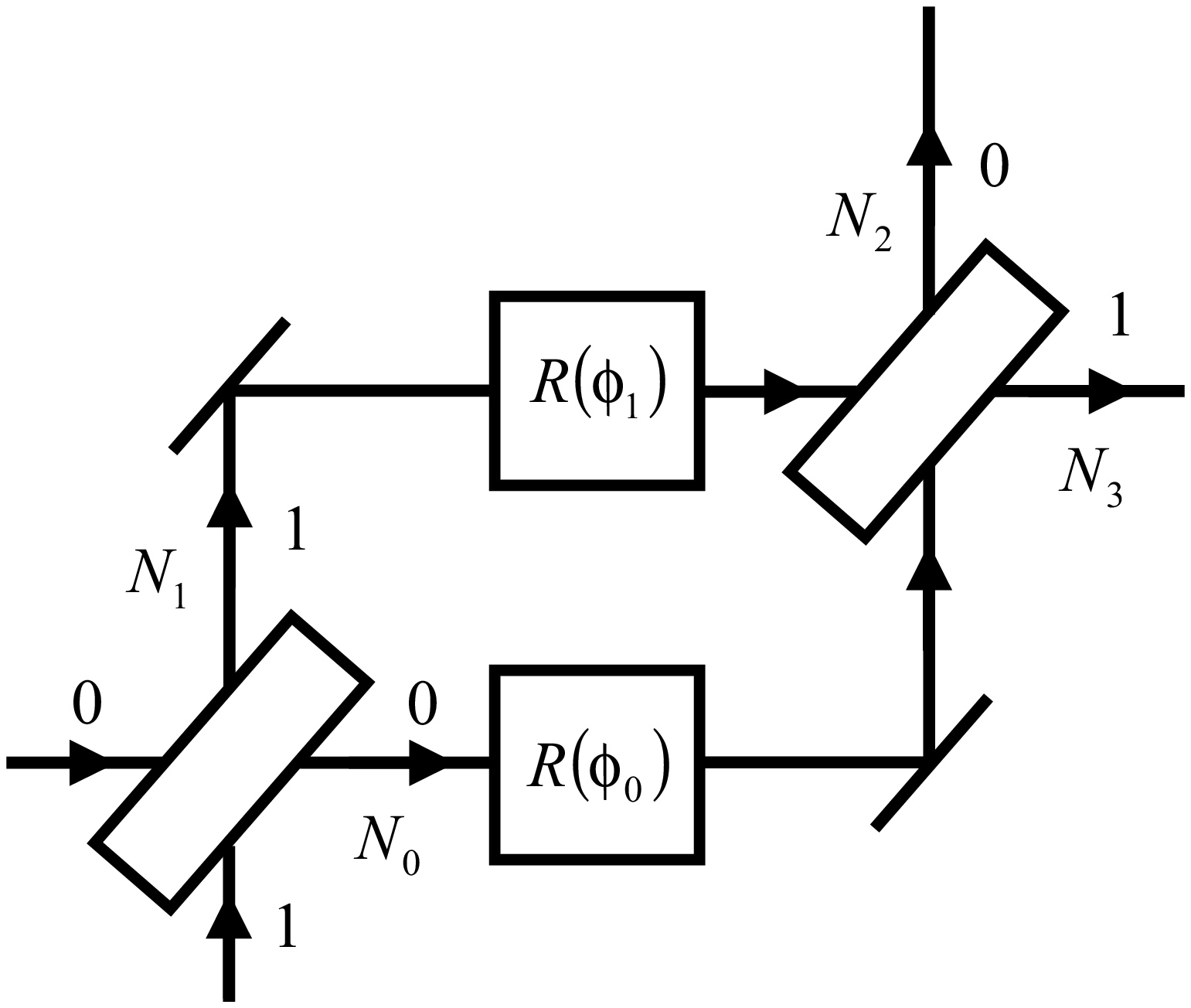}
\caption{
Diagram of a \DLM\ network that simulates a single-photon
Mach-Zehnder interferometer on an event-by-event basis~\cite{MZIdemo}.
The \DLM\ network consists of two BS devices (see Fig.~\ref{figbs})
and two passive devices $R(\phi_0)$ and $R(\phi_1)$ (see Eq.~(\ref{Rphi}))
that perform plane rotations by $\phi_0$ and $\phi_1$, respectively.
The number of events $N_i$ in channel $i=0,\ldots,3$
is proportional to the probability for finding a photon
on the corresponding arm of the interferometer.
}
\label{figmz}
\end{center}
\end{figure*}

\section{Single-Photon Beam Splitter}\label{BS}

In quantum theory,~\cite{QuantumTheory}
the presence of photons in the input modes 0 or 1 of a beam splitter is represented
by the complex-valued amplitudes ($a_0,a_1$)~\cite{BAYM74,GRAN86,RARI97}.
According to quantum theory, the complex-valued amplitudes ($b_0,b_1$)
of the photons in the output channels 0 and 1 of a beam splitter are given
by~\cite{BAYM74,GRAN86,RARI97}
\begin{eqnarray}
\left(
\begin{array}{c}
b_0\\
b_1
\end{array}
\right)
=
%\left(
%\begin{array}{c}
%a_0+ia_1\\
%a_1+ia_0
%\end{array}
%\right)
%=
\frac{1}{\sqrt{2}}
\left(
\begin{array}{cc}
1&i\\
i&1
\end{array}
\right)
\left(
\begin{array}{c}
a_0\\
a_1
\end{array}
\right),
\label{BS3}
\end{eqnarray}
Writing
$a_0=\sqrt{p_0} e^{i\psi_0}$ and
$a_1=\sqrt{1-p_0} e^{i\psi_1}$,
the probability to observe a photon in
output channel 0 (1) is given by
\begin{eqnarray}
|b_0|^2&=&\frac{1+\sqrt{p_0(1-p_0)}\sin(\psi_0-\psi_1)}{2},
\label{B0}
\\
|b_1|^2&=&\frac{1-\sqrt{p_0(1-p_0)}\sin(\psi_0-\psi_1)}{2}.
\end{eqnarray}
Here $\psi_0$ and $\psi_1$ represent the phases of the photons.
In a quantum theoretical description, this phase
is proportional to the length of the optical path that the photons
have travelled before they enter the beam splitter~\cite{BAYM74,GRAN86,RARI97}.

We now show that the \DLM-network shown in Fig.\ref{figbs}
behaves as if it is a single-photon beam splitter.
This network receives events at one of the two input channels.
There is a one-to-one relation between each input channel
and the corresponding input mode of the quantum mechanical description.
Each input event carries information in the form of a two-dimensional unit vector.
Either input channel 0 receives $(y_{0},y_{1})=(\cos\psi_0,\sin\psi_0)$
or input channel 1 receives $(y_{2},y_{3})=(\cos\psi_1,\sin\psi_1)$.
In terms of the single-photon experiments of Grangier et al.~\cite{GRAN86},
an event corresponds to the arrival of a photon at channel 0 (1) with
phase $\psi_0$ ($\psi_1$) of the beam splitter (see Fig.~\ref{figbs}).

The input message is fed into the \DLM-network described in Section~\ref{sec2}.
The purpose of \DLM\ 1 is to transform the information contained in
two-dimensional input vectors (of which only one is present for any given input event),
into a four-dimensional unit vector.
The internal vector ${\bf x}$ of \DLM\ 1 learns about
the amplitudes ($a_0,a_1$): In the stationary regime we have
\begin{eqnarray}
x_0&\approx&\sqrt{p_0} \cos{\psi_0}
,\nonumber \\
x_1&\approx&\sqrt{p_0} \sin{\psi_0}
,\nonumber \\
x_2&\approx&\sqrt{1-p_0} \cos{\psi_1}
,\nonumber \\
x_3&\approx&\sqrt{1-p_0} \sin{\psi_1}
.
\end{eqnarray}
The four-dimensional internal vector of this device
is split into two groups of two-dimensional vectors
$( x_{0}, x_{3})$ and $( x_{2},x_{1})$
and each of these two-dimensional vectors is rotated by $45^\circ$.
Put differently, the four-dimensional vector is rotated
once in the (1,4)-plane about $45^\circ$ and once in the (3,2) plane about $45^\circ$.
The order of the rotations is irrelevant.
Physically, this transformation corresponds to the reflection of the photons by $45^\circ$
at the beam splitter.
The resulting four-dimensional vector is then sent to the input of \DLM\ 2.
The internal vector ${\bf z}$ of \DLM\ 2 learns about
the amplitudes ($b_0,b_1$): In the stationary regime we have
\begin{eqnarray}
z_0&\approx&\sqrt{p_0} \cos{\psi_0}-\sqrt{1-p_0} \sin{\psi_1}
,\nonumber \\
z_1&\approx&\sqrt{p_0} \sin{\psi_0}+\sqrt{1-p_0} \cos{\psi_1}
,\nonumber \\
z_2&\approx&\sqrt{1-p_0} \cos{\psi_1}-\sqrt{p_0} \cos{\psi_0}
,\nonumber \\
z_3&\approx&\sqrt{1-p_0} \sin{\psi_1}+\sqrt{p_0} \cos{\psi_0}
\end{eqnarray}
\DLM\ 2 sends $(z_{0},z_{1})$ through output
channel 0 if it used rule $m=1,2,3,4$ (see Eq.~\Eq{HYP1}) to update its internal state.
Otherwise it sends $(z_{2},z_{3})$ through output channel 1.

In Fig.~\ref{one-bs} we present results of discrete-event simulations using
the \DLM\ network depicted in Fig.~\ref{figbs}.
We denote the number of 0 (1) events by $N_0$ ($N_1$) and
the total number of events by $N=N_0+N_1$.
The correspondence with the quantum system is clear:
the probability for a 0 event is given by
$|b_0|^2\approx N_0/N$, $y_{0}^\prime=\hbox{Re } b_0/|b_0|$ and $y_{1}^\prime=\hbox{Im } b_0/|b_0|$.
The probability for a 1 event is
$|b_1|^2\approx N_1/N$, $y_{2}^\prime=\hbox{Re } b_1/|b_1|$ and $y_{3}^\prime=\hbox{Im } b_1/|b_1|$.
Before the simulation starts,
the internal vectors of the \DLMS\ are given a random value (on the unit sphere).
Each data point represents 10000 events.
All these simulations were carried out with $\alpha=0.99$.
The simulation procedure itself consists of four steps:
\begin{enumerate}
\item Use two uniform random numbers in the range $[0,360]$
to generate $\psi_0$ and $\psi_1$.
\item For fixed values of $\psi_0$ and $\psi_1$, generate 10000 input events.
Input channel 0 receives $(y_{0},y_{1})=(\cos\psi_0,\sin\psi_0)$
with probability $p_0$.
Input channel 1 receives $(y_{2},y_{3})=(\cos\psi_1,\sin\psi_1)$ with probability
$p_1=1-p_0$.
\item Count the number of output events $N_0$ ($N_1$)
in channel 0 (1), see Fig.~\ref{figbs}.
\item Repeat steps 1 to 3.
For each pair ($\psi_0$,$\psi_1$), store the results for $N_0$ ($N_1$).
 \end{enumerate}
Plotting $N_0/(N_0+N_1)$ and $|b_0|^2$ as a function
of $\phi=\psi_0-\psi_1$ yields the results shown in Fig.~\ref{one-bs}.
Actually, there is no need to use random numbers to
generate $\psi_0$ and $\psi_1$.
In Fig.~\ref{one-bs}, we only used this random process to
show that the order in which we pick $\psi_0$ and $\psi_1$
is irrelevant.
Random processes enter in the procedure to generate the input data only.
The \DLM\ network processes the events sequentially and deterministically.
From Fig.~\ref{one-bs} it is clear that the output of the deterministic
\DLM-based beam splitter reproduces the probability distributions as obtained from
quantum theory~\cite{QuantumTheory}.

\begin{figure*}[t]
\begin{center}
\includegraphics[width=14cm]{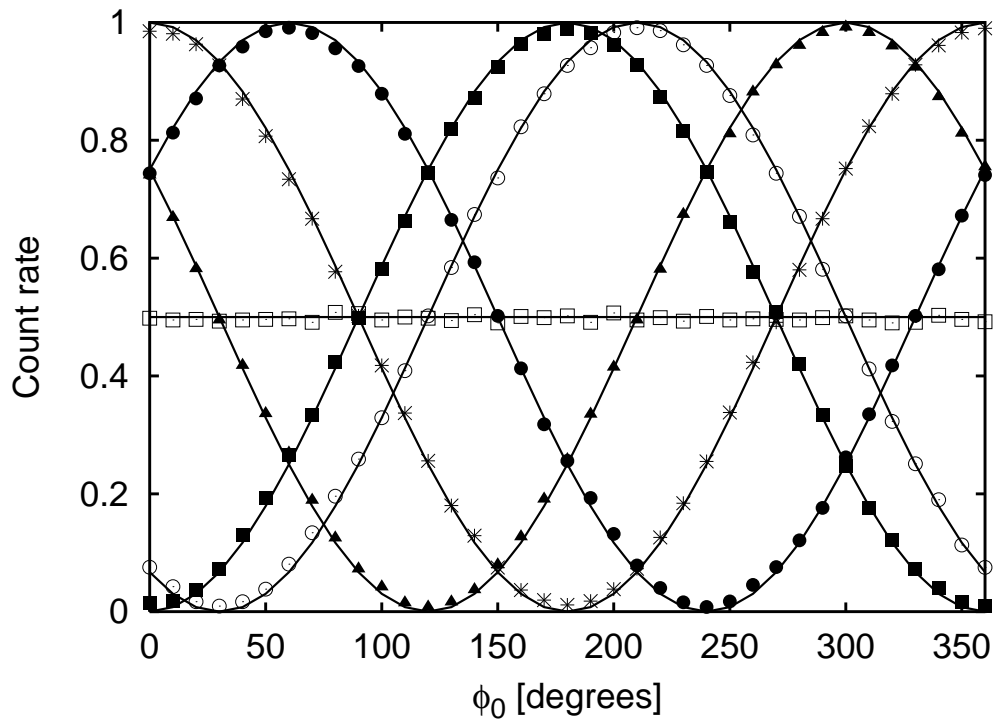}
\caption{
Simulation results for the \DLM-network shown in Fig.~\ref{figmz}.
Markers give the simulation results for the normalized intensities
as a function of $\phi_0$.
Open squares: $N_0/(N_0+N_1)$;
Solid squares: $N_2/(N_2+N_3)$ for $\phi_1=0$;
Open circles: $N_2/(N_2+N_3)$ for $\phi_1=30^\circ$;
Bullets: $N_2/(N_2+N_3)$ for $\phi_1=240^\circ$;
Asterisks: $N_3/(N_2+N_3)$ for  $\phi_1=0$;
Solid triangles: $N_3/(N_2+N_3)$ for $\phi_1=300^\circ$.
Lines represent the results of quantum theory.
}
\label{one-mz}
\end{center}
\end{figure*}

\section{Mach-Zehnder Interferometer}\label{MZI}

In quantum physics~\cite{QuantumTheory}, single-photon experiments with one beam splitter
provide direct evidence for the particle-like behavior of photons~\cite{GRAN86,HOME97}.
The wave mechanical character appears when one performs single-particle
interference experiments.
We now describe a \DLM\ network that displays the same
interference patterns as those observed in single-photon
Mach-Zehnder interferometer experiments~\cite{GRAN86}.
The schematic layout of the \DLM\ network is shown in Fig.~\ref{figmz}.
The network described in Section~\ref{BS} is used for the beam splitters.
The phase shift is taken care of by the
devices $R(\phi_0)$ and $R(\phi_1)$ (that do not contain \DLMS)
that perform plane rotations by $\phi_0$ and $\phi_1$ (see Eq.~(\ref{Rphi}), respectively.
Clearly there is a one-to-one correspondence between
the components of the \DLM\ network
and the elements of a physical Mach-Zehnder interferometer~\cite{BORN64,GRAN86}.

According to quantum theory~\cite{QuantumTheory}, the amplitudes ($b_0,b_1)$
of the photons in the output modes 0 ($N_2$) and 1 ($N_3$) of the
Mach-Zehnder interferometer are given by~\cite{BAYM74,GRAN86,RARI97}
\begin{eqnarray}
\left(
\begin{array}{c}
b_0\\
b_1
\end{array}
\right)
=U
\left(
\begin{array}{c}
a_0\\
a_1
\end{array}
\right)
,
\label{MZ1}
\end{eqnarray}
where
\begin{eqnarray}
U=
\frac{1}{2}
\left(
\begin{array}{cc}
1&i\\
i&1
\end{array}
\right)
\left(
\begin{array}{cc}
e^{i\phi_0}&0\\
0&e^{i\phi_1}
\end{array}
\right)
%\nonumber \\
%&\times&
\left(
\begin{array}{cc}
1&i\\
i&1
\end{array}
\right)
,
\label{MZ1a}
\end{eqnarray}
and $a_0$ ($a_1)$ denotes the amplitude
of the photons in the input channel 0 (1).
Note that in experiments it is impossible to
simultaneously measure ($N_0/(N_0+N_1)$, $N_1/(N_0+N_1)$)
and ($N_2/(N_0+N_1)$, $N_3/(N_0+N_1)$):
Photon detectors operate by absorbing photons.

In experiment~\cite{BORN64,GRAN86}, there are no photons
in input channel 1 of the first beam splitter, that is $a_1=0$.
From Eq.~(\ref{MZ1}), we see that the phase of the wave function describing
the photons in input channel 0 of the first beam splitter is irrelevant.
However, as the photons leave the first beam splitter the relation
between their phases is fixed. This relation can be changed
through the optical path length for reaching the second
beam splitter. A change of the optical path length
in channel 0 (1) results in a phase shift by $\phi_0$ ($\phi_1$).
If $a_1=0$, the output amplitudes of the Mach-Zehnder
interferometer are given by
\begin{eqnarray}
b_0&=&a_0 |a_0|^{-1}e^{i(\phi_0+\phi_1)/2}\sin\frac{\phi_0-\phi_1}{2},
\nonumber \\
b_1&=&a_0 |a_0|^{-1}e^{i(\phi_0+\phi_1)/2}\cos\frac{\phi_0-\phi_1}{2},
\label{MZ3}
\end{eqnarray}
from which we see that the probabilities $|b_0|^2$ and
$|b_1|^2$ depend on $\phi=\phi_0-\phi_1$ only.

In Fig.~\ref{one-mz} we present a representative selection of simulation results
for the Mach-Zehnder interferometer built from \DLMS.
We assume that input channel 0 receives
$(y_{0},y_{1})=(\cos\psi_0,\sin\psi_0)$ with probability
one and that input channel 1 receives no events.
This corresponds to $(a_0,a_1)=(\cos\psi_0+i\sin\psi_0,0)$ in Eq.~(\ref{MZ1}).
We use uniform random numbers to determine $\psi_0$.
As in the case of the beam splitter, we only use this random process to
show that the order in which we pick $\psi_0$ is irrelevant.
In all these simulations $\alpha=0.99$.
The data points are the simulation results for the
normalized intensity $N_i/(N_0+N_1)$ for i=0,2,3
as a function of $\phi=\phi_0-\phi_1$.
Each data point represents 10000 events ($N_0+N_1=N_2+N_3=10000$).
Initially the rotation angle $\phi_0=0$ and after each set of 10000 events, $\phi_0$
is increased by $10^\circ$.
Lines represent the corresponding results of quantum theory~\cite{QuantumTheory}.
From Fig.~\ref{one-mz} it is clear that the deterministic, event-based \DLM\ network
generates events with frequencies that are in excellent agreement with quantum theory.

\begin{figure*}[t]
\begin{center}
\includegraphics[width=14cm]{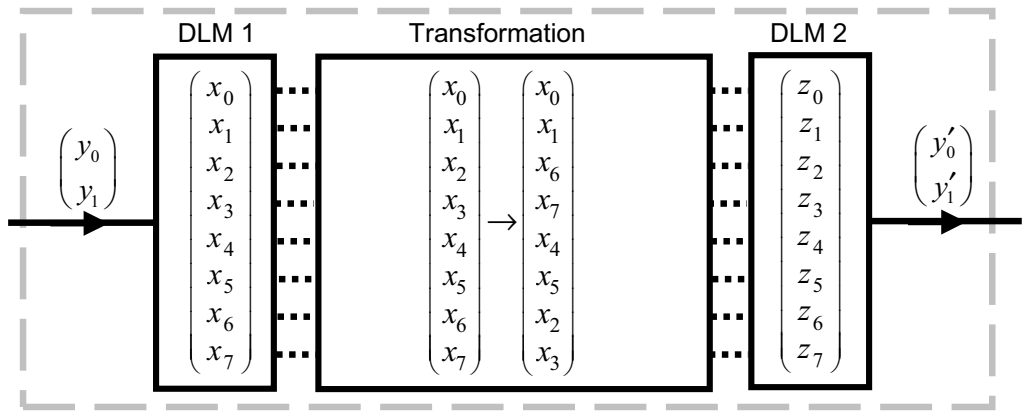}
\caption{
Diagram of a \DLM-based processor that simulates a CNOT gate
on an event-by-event basis.
}
\label{figcnot}
\end{center}
\end{figure*}

\begin{figure*}[t]
\begin{center}
\includegraphics[width=14cm]{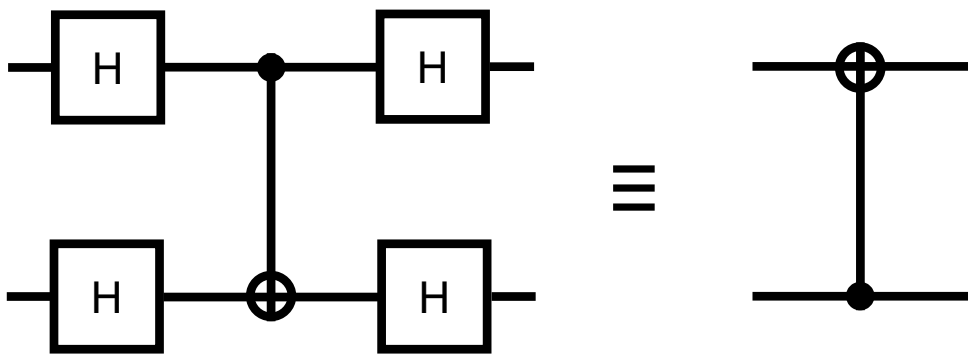}
\caption{
Quantum circuit representation of two equivalent
CNOT operations. The dot and the cross on the line
denote the control and target qubit, respectively.
The square boxes labeled $H$ represent Hadamard gates.
}
\label{figcnot2}
\end{center}
\end{figure*}

\section{Universal quantum computation}\label{CNOT}

It has been shown that an arbitrary unitary operation, that is,
the time evolution of a quantum system,
can be written as a sequence of single-qubit operations
and the controlled-NOT (CNOT) operation on two qubits~\cite{DIVI95a,NIEL00}.
Therefore, in principle, single-qubit operations and the CNOT operation
are sufficient to construct a universal quantum computer
or to simulate any quantum system~\cite{NIEL00}.
In this section, we present results of event-based simulations of
single qubit operations and a two-qubit quantum circuit containing
the CNOT operation to illustrate that \DLM-based networks can
be used to simulate universal quantum computers.

The state vector of a two-qubit system can be written
as~\cite{BAYM74,BALL03,NIEL00}
\begin{eqnarray}
\KET{\Phi}&=&
a_{0}\KET{0}_1\KET{0}_2+a_1\KET{1}_1\KET{0}_2
+a_2\KET{0}_1\KET{1}_2
\nonumber\\
&&+a_3\KET{1}_1\KET{1}_2
\nonumber\\
&=&
a_{0}\KET{00}+a_1\KET{01}+a_2\KET{10}+a_3\KET{11}
\nonumber\\
&=&
a_{0}\KET{0}+a_1\KET{1}+a_2\KET{2}+a_3\KET{3}
,
\label{CNOT1}
\end{eqnarray}
where $a_0,\ldots a_3$ are the amplitudes
of the four different states and
$\KET{0}_i$ and $\KET{1}_i$ represent the
0 and 1 state of the $i$-th qubit, respectively.
For convenvience, in the last line of Eq.(\ref{CNOT1}),
we represent the basis states of the two-qubit system
in decimal notation, that is $\KET{00}=\KET{0}$, $\KET{01}=\KET{1}$,
$\KET{10}=\KET{2}$, and $\KET{11}=\KET{3}$~\cite{NIEL00}.

\begin{table*}[t]
\caption{Simulation results for the \DLM-network shown in
Fig.~\ref{figcnot2}, demonstrating that the network reproduces
the results of the corresponding quantum circuit, that is, a CNOT operation
in which qubit 2 is the control qubit and qubit 1 is the target qubit~\cite{NIEL00}.
The first half of the events are discarded in the calculation
of the frequencies $f_i$ for observing an output event
of type $i=0,1,2,3$.
For 200 events or more, the difference between the event-based simulation results
and the corresponding quantum mechanical probabilities is less than 1\%.
}
\begin{tabular}{cccccccc}
Processor&Number of events&Qubit 1 &Qubit 2&$f_0$&$f_1$&$f_2$&$f_3$\\
\hline
\noalign{\vskip 4pt}
Deterministic&100&0&0&$0.98$&$0.00$&$0.00$&$0.02$\\ % 0.980E+00  0.000E+00  0.000E+00  0.200E-01
Deterministic&100&1&0&$0.20$&$0.74$&$0.01$&$0.04$\\ % 0.200E+00  0.200E-01  0.740E+00  0.400E-01
Deterministic&100&0&1&$0.16$&$0.04$&$0.00$&$0.80$\\ % 0.160E+00  0.000E+00  0.400E-01  0.800E+00
Deterministic&100&1&1&$0.16$&$0.04$&$0.72$&$0.08$\\ % 0.160E+00  0.720E+00  0.400E-01  0.800E-01
\noalign{\vskip 2pt}
\hline
\noalign{\vskip 2pt}
Deterministic&200&0&0&$1.00$&$0.00$&$0.00$&$0.00$\\
Deterministic&200&1&0&$0.00$&$1.00$&$0.00$&$0.00$\\
Deterministic&200&0&1&$0.00$&$0.00$&$0.00$&$1.00$\\
Deterministic&200&1&1&$0.01$&$0.00$&$0.99$&$0.00$\\
\hline
Stochastic&2000&0&0&$0.965$&$0.015$&$0.010$&$0.010$\\% 0.965E+00  0.100E-01  0.150E-01  0.100E-01
Stochastic&2000&1&0&$0.007$&$0.970$&$0.012$&$0.011$\\% 0.700E-02  0.120E-01  0.970E+00  0.110E-01
Stochastic&2000&0&1&$0.010$&$0.008$&$0.016$&$0.966$\\% 0.100E-01  0.160E-01  0.800E-02  0.966E+00
Stochastic&2000&1&1&$0.005$&$0.016$&$0.963$&$0.016$\\% 0.500E-02  0.963E+00  0.160E-01  0.160E-01
\hline
\end{tabular}
\label{CNOTdata}
\end{table*}

\begin{table*}[t]
\caption{Same as in Table ~\ref{CNOTdata} except that
the control parameter $\alpha=0.999$ instead of $\alpha=0.99$
and that ten times as many event were generated.
The results for the deterministic simulations are exact
within three-digit accuracy and have therefore been omitted.}
\begin{tabular}{cccccccc}
Processor&Number of events&Qubit 1 &Qubit 2&$f_0$&$f_1$&$f_2$&$f_3$\\
\hline
Stochastic&20000&0&0&$0.995$&$0.003$&$0.001$&$0.002$\\
Stochastic&20000&1&0&$0.002$&$0.995$&$0.003$&$0.001$\\
Stochastic&20000&0&1&$0.002$&$0.001$&$0.002$&$0.995$\\
Stochastic&20000&1&1&$0.001$&$0.002$&$0.997$&$0.001$\\
\hline
\end{tabular}
\label{CNOTdata2}
\end{table*}

\subsection{CNOT gate}

By definition, the CNOT gate flips the target qubit
if the control qubit is in the state $\KET{1}$~\cite{NIEL00}.
If we take qubit 1 (that is, the least significant bit in the binary notation of an integer)
as the control qubit, we have
\begin{eqnarray}
\CNOT\KET{\Phi}&=&
a_0\KET{0}_1\KET{0}_2+a_3\KET{1}_1\KET{0}_2+a_2\KET{0}_1\KET{1}_2
\nonumber\\
&&+a_1\KET{1}_1\KET{1}_2
\nonumber\\
&=&
a_0\KET{00}+a_3\KET{01}+a_2\KET{10}+a_1\KET{11}
\nonumber\\
&=&
a_0\KET{0}+a_3\KET{1}+a_2\KET{2}+a_1\KET{3}
.
\label{CNOT2}
\end{eqnarray}

The schematic diagram
of the \DLM-network that performs the CNOT operation on an
event-by-event (particle-by-particle) basis is shown in Fig.~\ref{figcnot}.
Conceptually the structure of this network
is the same as in the case of the Mach-Zehnder interferometer.
As input to the \DLM-network we now have four (0,1,2 or 3)
instead of two different types of events.
Each event carries a message consisting of two real numbers
${\bf y}=(y_{0},y_{1})$, corresponding to the quantum
mechanical amplitudes $a_0,\ldots a_3$.
The internal state of each \DLM\ is represented by a unit vector of eight real numbers
${\bf x}=(x_{0},\dots,x_{7})$ and there are
$16$ candidate update rules ($\{ j=0,\ldots7; s_j=\pm1\}$, see Eq.\Eq{HYP2})
to choose from.
The rule that is actually used is determined by minimizing the
cost function given by Eq.\Eq{HYP1}.
The transformation stage is extremely simple:
According to Eq.\Eq{CNOT2}, all it has to do is swap the two pairs
of elements ($x_2$,$x_3$) and ($x_6$,$x_7$).

Instead of presenting results that show that a \DLM-processor
correctly simulates the CNOT operation
on an event-by-event basis, we consider the more complicated
network of four Hadamard gates and one CNOT gate shown in Fig.~\ref{figcnot2}~\cite{NIEL00}.
Quantum mechanically, this network acts as a CNOT gate
in which the role of control- and target qubit have been interchanged~\cite{NIEL00}.
For the corresponding \DLM-network to work properly it is essential that
the event-based simulation mimics the quantum interference (generated by the Hadamard
gates) correctly.

\subsection{Hadamard operation}\label{HADA}

The Hadamard operation $H$ is the single-qubit operation
defined by~\cite{NIEL00}
\begin{equation}
H\equiv\frac{1}{\sqrt{2}}
\left(
\begin{array}{cc}
\phantom{-}1&\phantom{-}1 \\
\phantom{-}1&-1
\end{array}
\right)
.
\label{SPIN10}
\end{equation}
Disregarding phase factors, it performs the same operation as a beam splitter.

The structure of a \DLM-processor that performs
a general single-qubit operation is identical to the one shown in Fig.~\ref{figbs}.
The only difference is in the transformation stage.
To implement the Hadamard operation, we use the
transformation matrix $T$ (see Fig.~\ref{dlms})
\begin{equation}
\frac{1}{\sqrt{2}}
\left(
\begin{array}{cccc}
\phantom{-}1&\phantom{-}0 &\phantom{-}1 &\phantom{-}0 \\
\phantom{-}0&\phantom{-}1 &\phantom{-}0 &\phantom{-}1 \\
\phantom{-}1&\phantom{-}0 &-1 &\phantom{-}0 \\
\phantom{-}0&\phantom{-}1 &\phantom{-}0 &-1 \\
\end{array}
\right)
.
\label{HADA1}
\end{equation}

\subsection{Simulation results}

In Table~\ref{CNOTdata}, we present simulation results for the \DLM-network shown in
Fig.~\ref{figcnot2}.
Before the first simulation starts we
use uniform random numbers to initialize the
internal vectors of the \DLMS\ (ten vectors in total).
All these simulations were carried out with $\alpha=0.99$.
From Table~\ref{CNOTdata}, it is clear that, also for a modest number
of events, the network reproduces
the results of the corresponding quantum circuit, that is, a CNOT operation
in which qubit 2 is the control qubit and qubit 1 is the target qubit~\cite{NIEL00}.

As an illustration of the use of SLMs, we replace all the \DLM\ 2's
by SLMs in the \DLM\ implementation of the circuit shown in Fig.~\ref{figcnot2}
and repeat the simulations.
From Tables~\ref{CNOTdata} and ~\ref{CNOTdata2}, we conclude that the randomized
version generates the correct results but significantly more events
are needed to achieve similar accuracy as in the fully deterministic
simulation.

\subsection{Technical note}

All simulations that we presented in this section have been performed
for $\alpha=0.99$.
From the description of the learning process it is
clear that $\alpha$ controls the rate of learning or, equivalently,
the rate at which learned information can be forgotten.
Furthermore it is evident that the difference between
a constant input to a \DLM\ and the learned value of its
internal variable cannot be smaller than $1-\alpha$.
In other words, $\alpha$ also limits the precision with
which the internal variable can represent a sequence
of constant input values.
On the other hand, the number of events has to balance
the rate at which the \DLM\ can forget a learned input value.
The smaller $1-\alpha$ is, the larger the number of events
has to be for the \DLM\ to adapt to changes in the input data.

We use the example of this section to illustrate the effect of changing
$\alpha$ and the total number of events $N$.
In Table~\ref{CNOTdata2} we show the results of repeating the procedure used
to obtain the data shown in Table~\ref{CNOTdata} but
instead of $\alpha=0.99$ we used $\alpha=0.999$
and adjusted the number of events accordingly.
As expected, the difference between the simulation data and the results of
quantum theory decreases if $1-\alpha$ decreases and
the number of events increases accordingly.
Comparing Table~\ref{CNOTdata} with Table~\ref{CNOTdata2} it is clear
that the decrease of this difference is roughly proportional
to the inverse of the square root of the number of events.

\section{Discussion}
We have proposed a new procedure to construct algorithms
that can be used to simulate quantum processes
without solving the Schr{\"o}dinger equation.
There is a one-to-one correspondence
between the components of the network and the processing units
and the physical parts of the experimental setup.
Furthermore, only simple geometry is used to construct the simulation algorithm.
In this sense, the simulation approach we propose
satisfies Einstein's criteria of realism and causality~\cite{HOME97}.

An analogy may be helpful to understand the conceptual difference between
the conventional description of quantum theory and
the event-based approach proposed in this paper.
It is well known that an ensemble of simple, symmetric random walks may
be approximated by a diffusion equation (for vanishing lattice spacing
and time step). Also here we have two options. If we are interested in
individual events, we have no other choice than to simulate the discrete
random walk. However, if we want to study the behavior of many
random walkers, it is computationally much more efficient to solve
the corresponding diffusion problem.
The latter describes the outcome of (infinitly) many individual events
but does not provide information about individual events.
The random walk is the fundamental mechanism that gives rise to diffusion behavior.
In this sense, the \DLMS\ described in this paper may be regarded
as building blocks for a dynamic, deterministic, local and causal system that
generates individual events in such a manner that the collective behavior
of these events is described by quantum theory.

It may be of interest to compare our approach with
stochastic wavefunction methods~\cite{GISI95,DALI92,DUM92,GARD92,MOLM93,
GARR94,BREU95,BREU95a,EZAK95,BREU95b,BREU96,BRUN97,
MIYA98,BREU99,BREU03,BREU04}.
Instead of solving the equation of motion
of the density matrix, these methods solve stochastic
differential equations for an ensemble of independent realizations of pure states.
Typically, these methods are used to study
open quantum systems in which a small number
of degrees of freedom is coupled to a large reservoir.
An attempt to use a variant of the stochastic wavefunction method to
perform an event-by-event study of photon emission was reported in Ref.~\cite{MIYA98}.
In stochastic wavefunction methods, the wave function evolves
in time according to the time-dependent Schr\"odinger equation.
An uncorrelated random process interrupts this evolution to project
the wave function (that is, make a quantum jump) onto another normalized state.
This evolution of the wave function is similar to
the change of the internal state of a SLM if we consider one isolated event
that is processed by the transformation stage $T$ and a SLM (see Fig.~\ref{dlms}).
However, as a SLM is capable of learning from previous events,
the process of generating output events is non-Markovian,
this similarity being very superficial.
The fundamental difference between the two approaches can also be seen as follows.
In the stochastic wavefunction method, we can calculate
the time evolution of each member of the ensemble in parallel, at least in principle.
In the \DLM-approach, this is impossible: To exhibit quantum
mechanical behavior, it is imperative that the \DLM-network
processes events in a sequential manner.
In the fully deterministic \DLM-approach (that is, without the randomizing
feature of the SLM), there is no stochastic process at all.
Therefore, there also is no relation between the stochastic wavefunction method
and the deterministic, machine-learning approach discussed in this paper.

In conclusion, we have shown
that single-particle quantum interference and quantum computers
can be simulated on an event-by-event basis
using local and causal processes, without the need of
concepts such as wave functions or particle-wave duality.

\section*{Acknowledgment}
We thank Professor S. Miyashita for extensive discussions
and for critical readings of the manuscript.
We are grateful to Professors M. Imada and M. Suzuki for
many useful comments.

\raggedright


\begin{thebibliography}{99}
\bibitem{PerfectExperiments}
In this paper we disregard
limitations of real experiments such as detector efficiency,
imperfection of the source, biprism etc.
\bibitem{QuantumTheory}
We make a distinction between quantum theory and quantum physics.
We use the term {\sl quantum theory} when we refer to
the mathematical formalism, that is, the postulates
of quantum mechanics (with or without the wave function
collapse postulate)~\cite{BALL03} and the rules (algorithms) to compute the
wave function. We use the term {\sl quantum physics}
for microscopic, experimentally observable phenomena
that do not find an explanation within the mathematical
framework of classical mechanics.
\end{thebibliography}
\end{document}